\documentclass{article}
\usepackage{graphicx}        % For eps figures,newer & more powerfull
\usepackage{color}           % For color text: \color command
\usepackage{hyperref}
\usepackage{breakurl}        % For breaking URLs easily trough lines
\usepackage{natbib}
\newcommand{\etal}{{\it et al.}}

\begin{document}
\sloppy

%%%%%%%%%%%%%%%%%%%%%%%%%%%%%%%%%%%%%%%%%%%%%%%%%%%
\title{ Long-term variations  of the Sun's photospheric magnetic field }
\author{E.~S.~Vernova, M.~I.~Tyasto\\
IZMIRAN SPb Filial, St. Petersburg, Russia\\ 
elenavernova96@gmail.com\\
D.~G.~Baranov\\
Ioffe Institute, St. Petersburg, Russia
}
\date{}
\maketitle
%%%%%%%%%%%%%%%%%%%%%%%%%%%%%%%%%%%%%%%%%%%%%%%%%%%
 %%% Abstract
%\begin{abstract}
Variations of the weak magnetic fields of the photosphere with periods of the order of the solar magnetic cycle 
were  investigated. Synoptic maps of the photospheric magnetic field produced 
by NSO Kitt Peak for the period from 1978 to 2016 were used as initial data. 
In order to study weak magnetic fields, the saturation threshold for synoptic maps was set at 5 G. 
On the base of transformed synoptic maps the time-latitude chart was built. 18 profiles of the magnetic field  
evenly distributed along the sine of latitude from the north to the south pole were selected in the diagram for the 
further analysis. Time dependencies were averaged by sliding smoothing over 21 Carrington rotations. 
The approximation of averaged time dependencies by the sinusoidal function made it possible to distinguish in 
weak magnetic fields a cyclic component with a period of about 22 years (the period of the Hale magnetic cycle). 
The dependence of 22-year variation on latitude was studied. In addition to the well-known 22-year change in the 
near-polar field, similar variations were found for the fields at all latitudes. The exception was latitudes $26^\circ$ 
and $33^\circ$ in the northern and $26^\circ$ in the southern hemisphere. These mid-latitude intervals were characterized 
by a predominance of short-period variations. The amplitude of the long-term variation decreased from the poles to the equator, 
with the period of variation remaining almost constant (T = 22.3 years). 
%\end{abstract}
%%%%%%%%%%%%%%%%%%%%%%%%%%%%
\section{Introduction}
During the solar cycle, radical changes in the magnetic fields of the Sun occur, and, in particular, the change of magnetic field distribution over the surface of the Sun takes place. The rearrangement of the distribution reflects processes that occur not only on the visible surface of the Sun, but also in the deeper layers, under the photosphere. Usually while developing the model of the 11-year cycle of solar activity, first of all,  the evolution of the  sunspots is considered. Sunspots contain the strongest magnetic fields, making it easier to observe the change in the distribution of activity over the surface of the Sun. The pattern of the sunspot distribution in the time-latitude diagram is clearly visible in the form of Maunder butterflies.

Cyclic changes in the distribution of solar magnetic field depend both on the intensity of the specific field groups and on the polarity of magnetic field. According to Hale's law, polar magnetic field changes its sign near the maximum of solar activity, while the sign of the leading and following sunspots changes during the period of solar minimum. Thus, the complete magnetic cycle of the Sun consists of two 11-year cycles and is about 22 years (Hale cycle).

In the formation of the solar cycle, the transfer of magnetic fields by the flows in the photosphere (surges),  plays an important role. Studies of  the surges established a close relationship between polar field reversal and the transfer of magnetic fields to the pole by rush-to-the-pole (RTTP) flows. This phenomenon was studied in the green coronal line \citep{alt}, it was also repeatedly observed in high-latitude prominence eruptions  \citep{gopa}. The role of RTTP flows in photospheric magnetic field transfer has been investigated in studies \citep{petr, mord}. Drifting from latitudes $\sim 40^\circ$  to poles, these streams are the product of following sunspots decay and always have the following sunspot polarity. The length of RTTP flows is about three years, their arrival at the poles of the Sun causes a change in the sign of the polar field. The lifetime of RTTP always falls on the period of the solar activity maximum, while the arrival at the pole coincides with the polarity change of the polar field. 
The migration of magnetic fluxes from active regions to the poles and their involvement in polar field inversion was examined in \citep{sun} for the Solar Cycle 24. A statistical method for analyzing the transfer of fields to the poles during cycles 21--24 has been proposed in \citep{wang}.

\citet{vecc} and \citet{ulri} described a new phenomenon in the distribution of magnetic fields, which consists in the appearance of wave-like structures with a period of about 2 years. 
In \citep{vecc}  for each heliographic longitude the radial component of the field measured at NSO Kitt Peak was decomposed into internal modes with the help of the Empirical Mode Decomposition  method. As a result, the migration of magnetic fields to the poles during the period of maximum and decline of the solar cycle was discovered, which the authors associated with the manifestation of quasi-biennial variations. This phenomenon has been investigated in detail in \citep{ulri}, where the term ``ripples'' has been proposed for such magnetic fluxes. In \citep{ulri} these ripples were detected by differentiation of the time-latitude diagram constructed on the base of the Mount Wilson Observatory data. In the differentiated diagram, ripples were seen at any level of solar activity. Another characteristic parameter considered in \citep{ulri} was the deviation from the trend.  This parameter also shows the alternation of ripples of opposite signs which is present regardless of the level of solar activity. Using different data and methods of their treatment \citep{vecc} and  \citep{ulri} obtained a number of similar results, indicating the appearance at low latitudes and movement to the poles of wave structures with a period close to the period of quasi-biennial variations.

Patterns of distribution of weak magnetic fields are not as well understood as distribution of strong magnetic fields, although it is the weak fields that occupy most of the Sun's surface. According to our estimates \citep{vern1} performed with the NSO Kitt Peak data for the period 1978--2016, magnetic fields with intensity $| B | \leq 10$ G occupied more than $82\%$ of the surface  of the Sun. The spatio-temporal evolution of weak magnetic fields has been examined in \citep{geta1, geta2, murs}. The distribution of weak photospheric fields was studied in \citep{geta1} using different data sets. In \citep{geta2} the asymmetry of weak magnetic fields was considered separately for each of the hemispheres, and it was shown that shifts in the northern and southern hemispheres are generally opposite.

The purpose of this work is to study the distribution of weak magnetic fields of the positive and negative polarities over the surface of the Sun. This work continues the study of cyclic processes in the magnetic field of the photosphere that we began in \citep{vern1}. The main focus is now on long-period changes in the magnetic field, which are considered over almost four solar cycles.

\section{Data and Method}
When studying the magnetic fields of the photosphere, the method of synoptic maps and time-latitude diagrams has become widespread. In our study synoptic maps produced at the National Solar Observatory (NSO Kitt Peak) were used. Measurements of the Sun's magnetic field were made by the Kitt Peak Vacuum Telescope (KPVT) in 1976--2003 (ftp:
//nispdata.nso.edu/kpvt/synoptic/mag/) and Synoptic Optical Long-term Investigations of the Sun (SOLIS equipment) in 2003--2016 (https://magmap.nso.edu/solis/archive.html). In the period before 1978 the number of missing data  was  significant. Taking into account only the period when observations became quite regular (1978--2016), we included  in our analysis  521 synoptic maps of the Sun's magnetic field. Each map corresponded to one of Carrington rotations and contained 180x360 pixels (one map for  each Carrington rotation) of magnetic field values in Gauss units with a resolution 
of $1^\circ$ longitude and 180 steps along the sine of latitude.
The synoptic maps are subject to both systematic and random errors. Due to the tilt of the Sun's axis of rotation part of the time it becomes impossible to observe the near-polar regions. Filling the lacunae in the data by extrapolating the measurements gives less reliable results. The source of error in synoptic maps are random fluctuations (noise), which in the polar regions are about 2 G according to estimates in \citep{harv}. However, as time-latitude diagrams are  obtained by averaging of the 360 field values  over longitude, the  random error is reduced down to  a value of about 0.1 G.

Synoptic maps averaged over longitude taking into account the sign of the magnetic field were used for the construction of the time-latitude diagram. Since we were interested in the properties of the weak fields, we limited the influence of strong fields, thereby emphasizing the contribution of weak fields to the picture of the magnetic field distribution over the surface of the Sun. To this end, a saturation threshold of 5 G was set for each synoptic map. As a result, only fields with modulo less than 5 G were left unchanged in each synoptic map, while larger or smaller fields were replaced by the corresponding threshold values of $+ 5$~G or  $-5$~G. Maps thus converted were used to build the time-latitude diagram.

Our work focuses mainly on the study of the distribution of weak magnetic fields along the latitude and its rearrangement in time. Weak fields make up a significant part of the Sun's magnetic field: 
fields with intensity $|B|< 5$~G occupy about $65\%$ of the solar surface in the period of 1978--2016 \citep{vern1}. 
%%%%%%%%%%%%%%%%%%%%%%%%%%%%%%%%%%%%%%%%%%%%%%%%%%%%%%%%%%Figure1
 \begin{figure}[ht]
   \centerline{\includegraphics[width=0.95\textwidth,clip=]{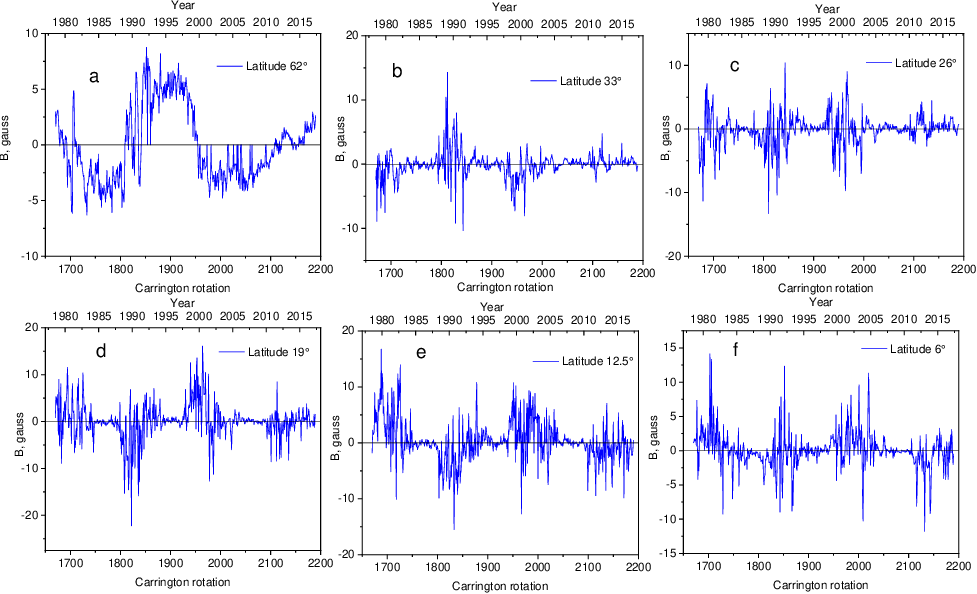}
              }          
							\caption{
							Time dependencies of the magnetic field  where all strength values $B$ were taken into account. Graphs were obtained at fixed latitudes of the time-latitude diagram, 
							built on the base of synoptic maps for 1670--2190 Carrington rotations. Following latitudes of the northern hemisphere were selected: a) $62^\circ$; b) $33^\circ$; c) $26^\circ$; 
							d) $19^\circ$; e) $12.5^\circ$; f) $6^\circ$.
                      }
      \label{rows}                                        
   \end{figure}
%%%%%%%%%%%%%%%%%%%%%%%%%%%%%%%%%%%%%%%%%%
The choice of the magnetic field intensity $ | B |= 5 $ G as a threshold value is connected with the following feature of the photospheric magnetic field. The flux of fields with $|B| <5$ G varies in antiphase with the flux of stronger fields which follows the 11-year solar cycle \citep{vern2}. A similar result was obtained in \citep{jin}, where it was shown that magnetic structures with low flux values change in antiphase with the solar cycle. It should be noted that in our work strong fields are not completely excluded from the analysis: in synoptic maps the intensities  $|B|> 5$ G were replaced with threshold values  of  $\pm 5$ G
for positive and negative fields, respectively, and thus made significant addition to the weak fields.
%%%%%%%%%%%%%%%%%%%%%%%%%%%%%%%%%%%%% Figure 2
 \begin{figure}[ht]
   \centerline{\includegraphics[width=0.95\textwidth,clip=]{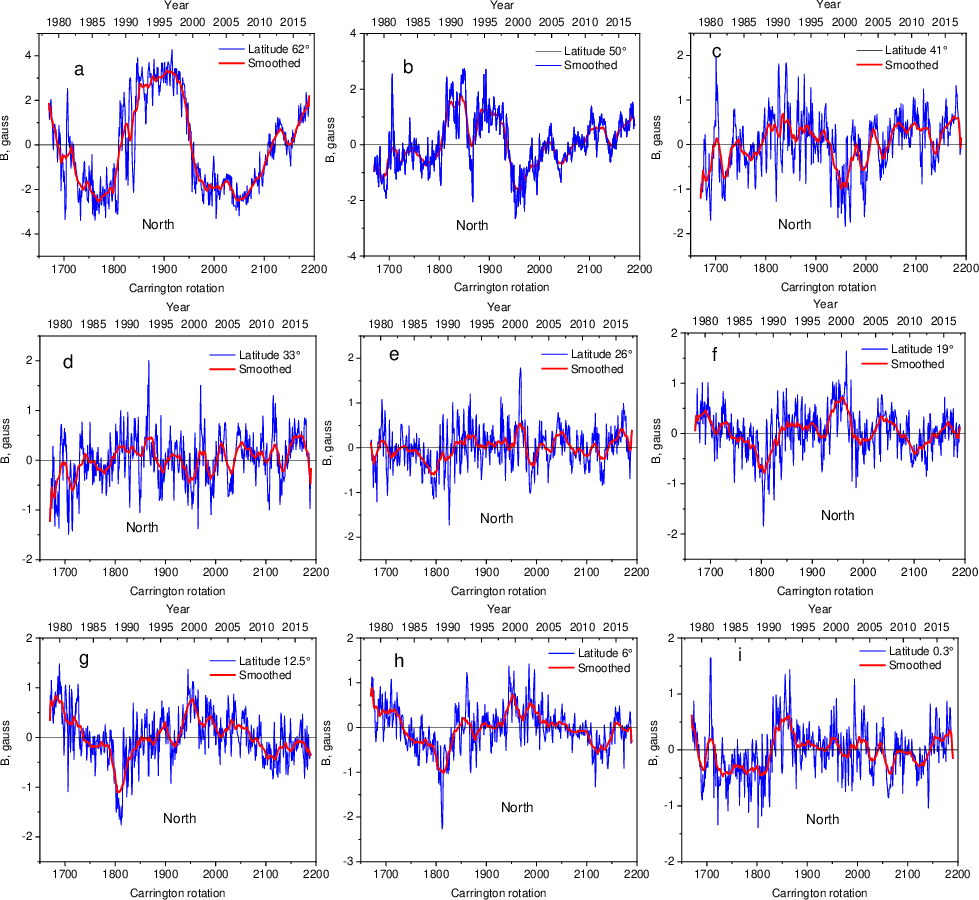}
              }          
							\caption{Time change of the weak magnetic field (northern hemisphere).   Dependencies for the selected latitudes were obtained from the time-latitude diagram, 
							which was built using transformed synoptic maps. In each synoptic map, the fields $|B|\leq 5$ G were left unchanged, while pixels with $ B > 5 $ G were filled with the threshold 
							value $B=5$ G. Correspondingly, values $ B < -5 $ G were replaced by the lower threshold $B=-5$ G. Magnetic field change was plotted for the latitudes of the northern 
							hemisphere: a) $62^\circ$; b) $51^\circ$; c) $41^\circ$; d) $33^\circ$; e)~$26^\circ$; f) $19^\circ$; g) $12.5^\circ$; h) $6^\circ$; i) $0.3^\circ$. 
							Blue line -- primary data, red line -- data averaged by sliding smoothing over 21 Carrington rotations.							
                       }
      \label{cols}                                        
   \end{figure}
	%%%%%%%%%%%%%%%%%%%%%%%%%%%%%%%%%%%%%%%%%%%%%%%%%%%%

\section{Results and Discussion}
\label{work results}
Although our work is devoted to the distribution of weak magnetic fields over the surface of the Sun, we considered for comparison also the time-latitude diagram containing all values of the magnetic field strength. In this diagram some fixed latitudes were selected and temporal changes in magnetic fields at these latitudes were examined.

Figure~\ref{rows} presents the changes in the magnetic field in the northern hemisphere along the latitudes: $62^\circ$, $33^\circ$, $26^\circ$, $19^\circ$, $12.5^\circ$  and $6^\circ$. Figure~\ref{rows}a clearly shows a 22-year magnetic cycle for the near-polar region ($62^\circ$). 
At other latitudes, a different picture is observed: sharp fluctuations in the magnetic field during years of high solar activity and almost complete absence of variations  the rest of the time. Similar results were obtained in \citep{vecc}, where the change of the magnetic field along three latitudes: $60^\circ$, $25^\circ$ and $5^\circ$ was examined.

Now let us consider  the time-latitude diagram obtained from synoptic maps with saturation at the threshold  value of 5 G. The time-latitude diagram was used further for constructing a set of magnetic field temporal dependencies at 18 different latitudes, which were distributed evenly along the sine of the latitude from the north to the south pole.

%%%%%%%%%%%%%%%%%%%%%%%%%%%%%%%%%%%%%%%%%%% Figure 3

 \begin{figure}[t]
   \centerline{\includegraphics[width=0.95\textwidth,clip=]{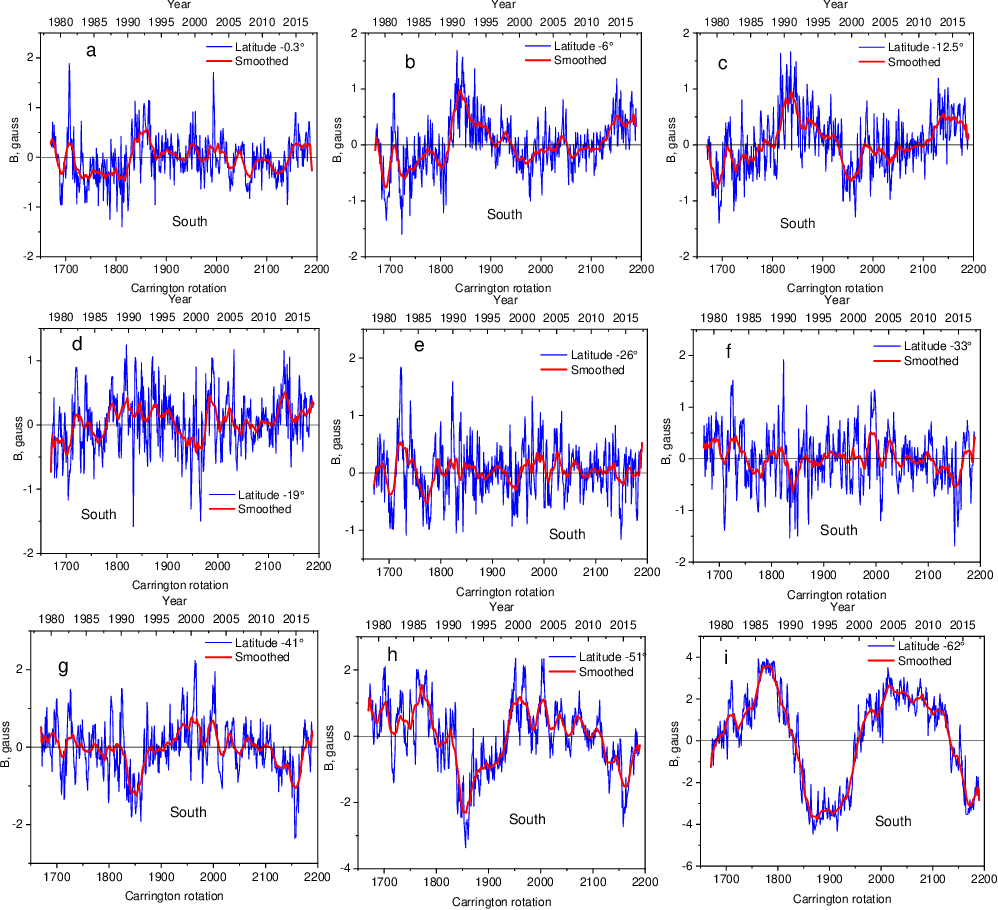} 
              }          
							\caption{
							The same as in Figure~\ref{cols} for the latitudes of the southern hemisphere: a)~$-0.3^\circ$; b)~$-6^\circ$; c)~$ -12.5^\circ$; d)~$-19^\circ$; e)~$-26^\circ$; f)~$-33^\circ$; 
							g)~$-41^\circ$; h)~$-51^\circ$; i)~$-62^\circ $.  Blue line -- primary data, red line -- data averaged by sliding smoothing  over 21 Carrington rotations.
                      }
      \label{tabs}                                        
   \end{figure}

Figure~\ref{cols}  (northern hemisphere) and Figure~\ref{tabs}  (southern hemisphere) show the change in magnetic field over four solar cycles at fixed latitudes as a function of time. The sign of the field varies irregularly from one rotation to the other throughout the  time in question. Especially often the changes in polarity occur at latitudinal profiles  $26^\circ$ and $33^\circ$.  Despite the strong fluctuations of the primary data in Figures~\ref{cols} and \ref{tabs}, it can be seen that in the near-polar regions (Figures ~\ref{cols}a and \ref{tabs}i) the field sign changes cyclically with a period of about 22 years (the magnetic cycle of the Sun).
%%%%%%%%%%%%%%%%%%%%%%%%%%%%%%%%%%%%%%%%%%% Figure 4

 \begin{figure}[t]
   \centerline{\includegraphics[width=0.95\textwidth,clip=]{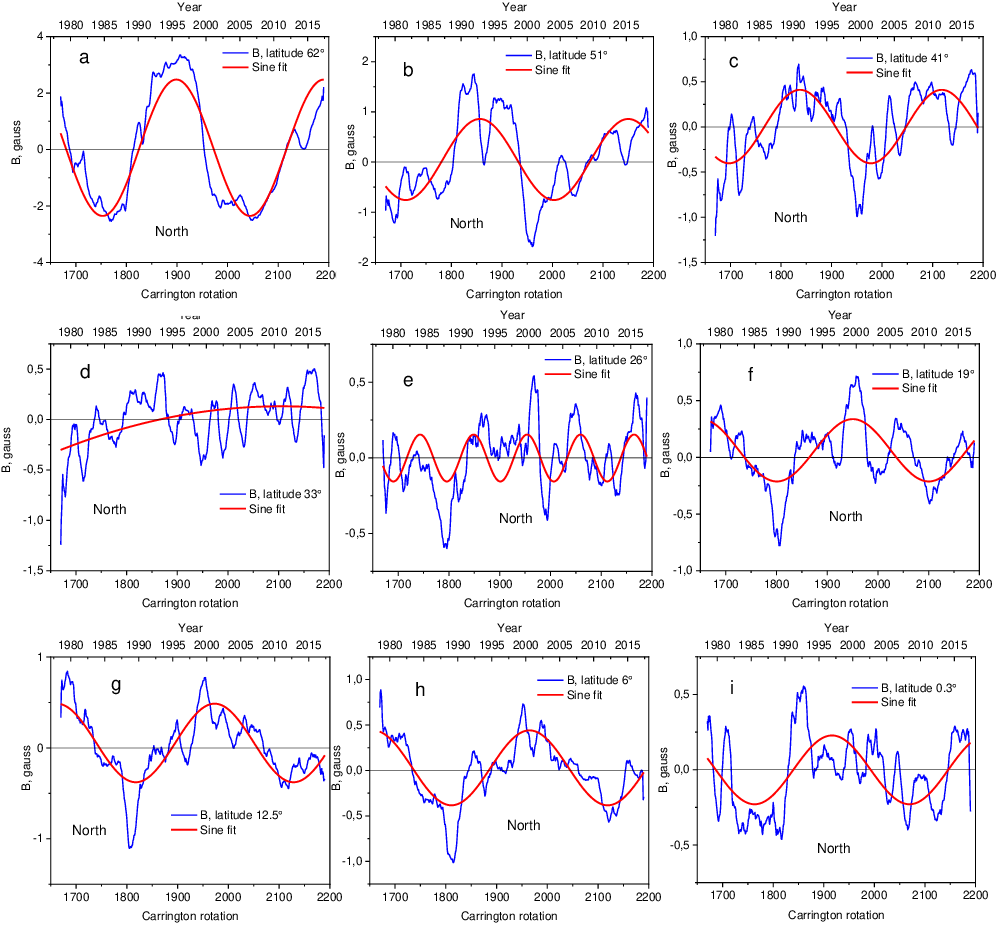} 
              }          
							\caption{
							Northern  hemisphere.  Approximation by the sinusoidal function (red line) of the smoothed   magnetic  field  data (blue line) at latitudes:  
							a) $62^\circ$; b) $51^\circ$; c) $41^\circ$; d) $33^\circ$; e) $26^\circ$; f) $19^\circ$; g) $12.5^\circ$; h) $6^\circ$; i) $0.3^\circ$. 
							There is no long-period variation at latitudes $26^\circ$ 
							 and $33^\circ$ 	(Figures~\ref{nappr}d,e).
							%(Figure~\{}d and Figure~\{}e).
                      }
      \label{nappr}                                        
   \end{figure}

When comparing the time development of the strong fields (Figure~\ref{rows}c) with the weak fields (Figure~\ref{cols}e) (both figures show the latitude $26^\circ$ of the northern hemisphere), one can see their significantly different behavior. While Figure~\ref{rows}c shows the presence of an 11-year periodicity, Figure~\ref{cols}e displays chaotic polarity changes that do not obey any pattern resembling an 11-year cycle. Thus, the distribution of weak magnetic fields has specific features, the study of which allows  to obtain new information complementing  the overall picture of the distribution of photospheric fields.
%%%%%%%%%%%%%%%%%%%%%%%%%%%%%%%%%%%%%%%%%%% Figure 5

 \begin{figure}[t]
   \centerline{\includegraphics[width=0.95\textwidth,clip=]{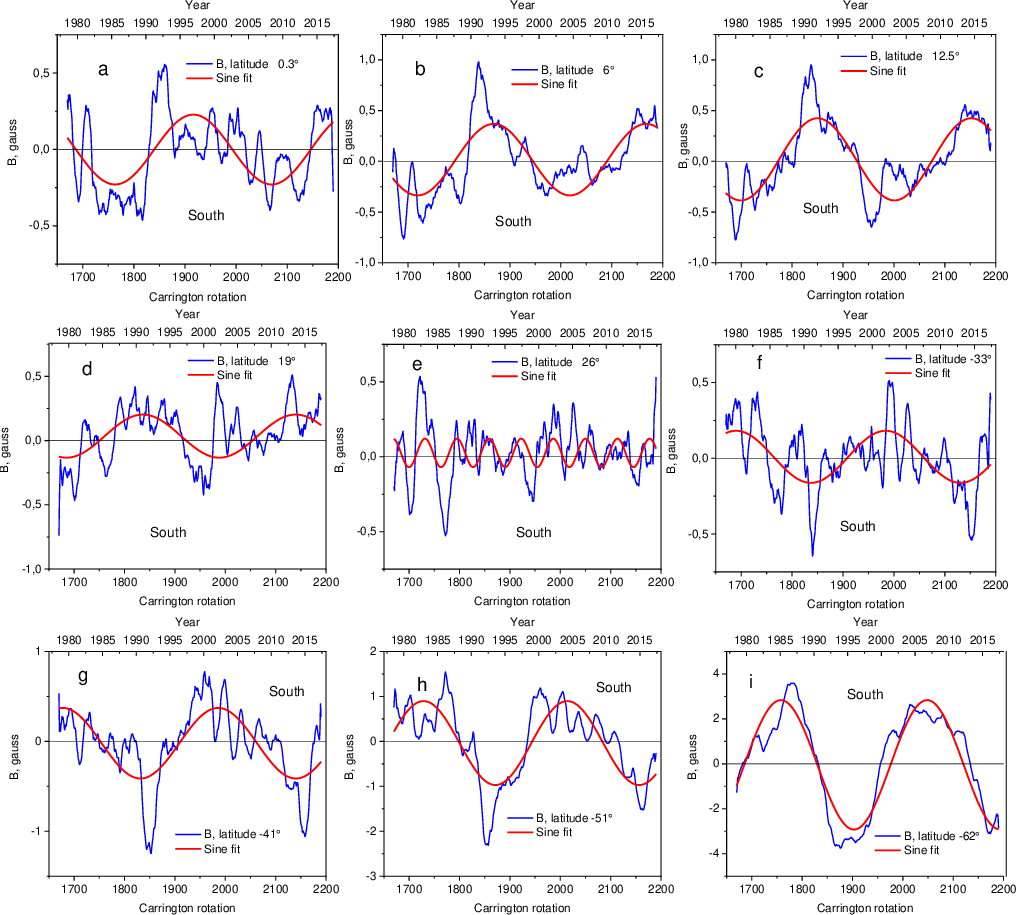} 
              }          
							\caption{
							Southern  hemisphere.  Approximation by the sinusoidal  function (red line) of the smoothed magnetic  field data (blue line) at latitudes: 
							a)~$-0.3^\circ$; b)~$-6^\circ$;  c)~$-12.5^\circ$; d)~$-19^\circ$;  e)~$-26^\circ$; f)~$-33^\circ$; 
							g)~$-41^\circ$; h)~$-51^\circ$; i)~$-62^\circ$.  
                      }
      \label{sappr}                                        
   \end{figure}

To study the long-period variations the presence of the variations with the shorter periods should be taken into account. Cyclic changes of magnetic fields with periods near to quasi-biennial variations were observed by \citep{vecc}, while \citep{ulri} described variations with periods ranging from 0.8 to 2 years. In \citep{vern1}, the existence of variations in the magnetic field with periods of 1.6--1.8 years was shown. To suppress  the influence of these variations and highlight variations with longer periods, the primary data were processed in this study by the sliding  smoothing over 21 Carrington rotations. Such smoothing reduces the amplitude of variations by 5--10 times as compared with periods of less than 2 years.

Smoothed time dependencies (red curves) are shown in Figure~\ref{cols} for the  latitudes from $62^\circ$ of the northern hemisphere to the equator; Figure~\ref{tabs} shows the same set of graphs for the southern hemisphere. A certain regular pattern can be seen in these graphs with the dipole component of the magnetic field dominating in the near-polar regions.

The strictly ordered structure of the near-polar regions (Figures~\ref{cols}a and \ref{tabs}i) is gradually distorted, and short-period variations are observed at latitudes $26^\circ$ and $33^\circ$ of the northern hemisphere and $26^\circ$ of the southern hemisphere instead of a large-scale time patterns. Surprisingly, while moving  farther  towards the equator, the dipole nature of the magnetic field manifests itself  again (see Figures~\ref{cols}g and \ref{tabs}c).

To  separate the long-term variations of the magnetic field, smoothed time relationships (Figure~\ref{nappr} and \ref{sappr}) 
were approximated by the sinusoidal function of the form:
%%%%%%%%%%%%%%%%%%%%%%%%%% Equation 1
\begin{equation}\label{fsin}
y=y_0+A \sin(\pi(x-x_c)/w),
\end{equation}                                                                 
where $A$ is the amplitude, $w$ is the half-period, $x_c$ is the phase of variation, and $y_0$ is the constant level.

\begin{figure}[t]
   \centerline{\includegraphics[width=0.95\textwidth,clip=]{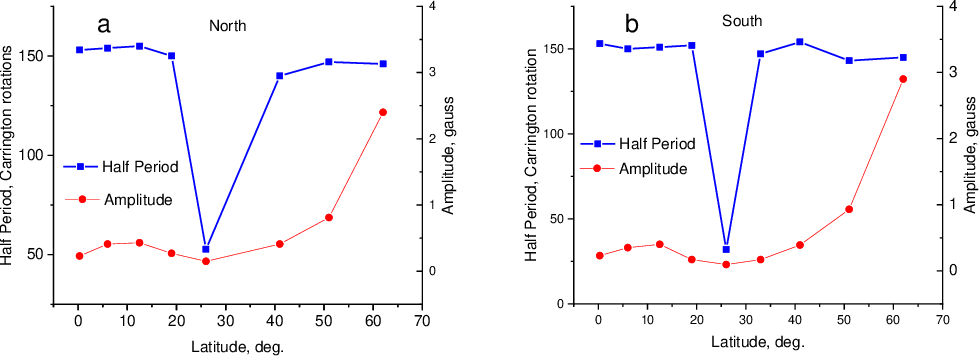} 
              }          
							\caption{Magnetic field variation parameters: (a) northern hemisphere; and (b)~the southern hemisphere. The amplitude of the variation is represented 
							by a red line, the half-period of the variation is represented by a blue line. A sharp drop of the period values occurs at latitude $+26^\circ$ of the 
							northern hemisphere and at $-26^\circ$ of the southern hemisphere. At these latitudes, instead of long-period variation, a variation with a shorter period is observed.
                      }
	\label{TandAmp}

   \end{figure}

\begin{figure}[t]
   \centerline{\includegraphics[width=0.8\textwidth,clip=]{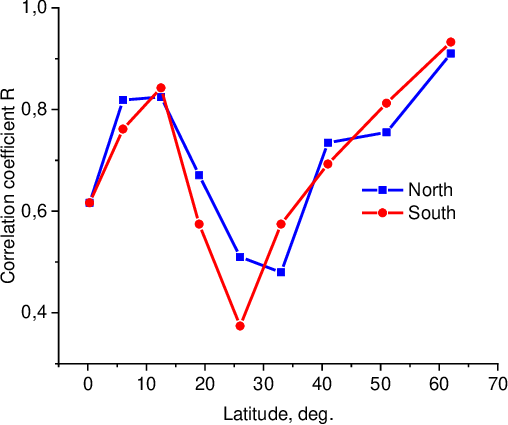} 
              }          
							\caption{The spread of experimental points relative to the approximating sinusoid evaluated by the correlation coefficient $R$ for the considered set of latitudes 
							from the equator to $+62^\circ$ (northern hemisphere, blue line) and from the equator to $-62^\circ$  (southern hemisphere, red line).
                      }
	\label{corr}

   \end{figure}

Approximation was performed using the NLCF (Non Linear Curve Fitting) standard procedure of the Origin graphics and computational package. The procedure uses the iterative Levenberg--Marquardt method (L-M algorithm). As a rule, the process was completed after 9--10 iterations, when the difference between the reduced chi-square values of two consecutive iterations became less than the admissible value of $1\times10^{-9}$. The fitting results for the set of latitudes from $+62^\circ$ to $-62^\circ$ are shown in Table~\Ref{AppRes} and in Figures~\ref{TandAmp} and \ref{corr}. In Table~\Ref{AppRes} the near-equatorial latitudes of the northern and southern hemispheres are averaged and considered as one resulting value. For the latitude $33^\circ$  of the northern hemisphere (Figure~\ref{nappr}), the period of variation could not be determined, and the calculation results are not included in Table~\Ref{AppRes}.

 \begin{table}
\caption{Magnetic field approximation at selected latitudes from $+62^\circ$  to $-62^\circ$.
}

\label{AppRes}
\begin{tabular}{ccccc}     % define the column alignment
                           % l: left,c: center,r: right
  \hline                   % horizontal line
%%Feature & RTTP & Ripples\\
Lat &$+62^\circ$               &$+51^\circ$              	&$+41^\circ$                  &$+26^\circ$\\
  \hline
y0	&$0.063 \pm 0.036$	&$0.051  \pm  0.023$	 &$0.004  \pm  0.01$	&$-0.001  \pm  0.01$\\
xc	&$156.75 \pm 1.08$	&$114.0 \pm 2.90$	&$98.24 \pm 3.06$	&$47.89 \pm 2.23$\\
w	&$145.84 \pm 0.91$	&$146.82 \pm 2.11$	&$140.36 \pm 1.88$	&$52.69 \pm 0.45$\\
A	&$2.42 \pm 0.049$	&$0.81 \pm 0.03$	&$0.406 \pm 0.017$	&$0.15 \pm 0.012$\\
COD	&  0.83	& 0.57	& 0.53	& 0.26\\
R	& 0.91	& 0.76	& 0.73	& 0.51\\
\hline
Lat &$+19^\circ$               &$+12.5^\circ$              	&$+6^\circ$                  &$0^\circ$\\
\hline
y0	&$0.062 \pm 0.01$	&$0.055 \pm 0.01$	&$0.029 \pm 0.009$	&$0.0013 \pm 0.01$\\
xc	&$-92.89 \pm 5.57$	&$-84.55 \pm 3.89$	&$-88.94 \pm 3.80$	&$-141.3 \pm 6.80$\\
w	&$149.73 \pm 2.14$	&$155.39 \pm 1.62$	&$154.0 \pm 1.54$	&$153.11 \pm 2.42$\\
A	&$0.27 \pm 0.013$	&$0.43 \pm 0.013$	&$0.412 \pm 0.013$	&$0.24 \pm 0.013$\\
COD	&0.45	&0.68	&0.67	&0.40\\
R	&0.67	&0.82	&0.82	&0.63\\
\hline
Lat &$-6^\circ$               &$-12.5^\circ$              	&$-19^\circ$                  &$-26^\circ$\\
\hline
y0	&$ 0.017 \pm 0.01$	&$0.02   \pm 0.009$	&$0.035  \pm 0.008$	  &$0.026 \pm 0.007$\\
xc	&$124.51 \pm 2.70$	&$105.61 \pm 2.31$	&$91.65  \pm 5.19$	  &$46.56 \pm 1.95$\\
w	  &$149.59 \pm 2.10$	&$150.91 \pm 1.67$	&$151.66 \pm 3.55$	  &$31.84 \pm 0.24$\\
A	  &$  0.35 \pm 0.013$	&$0.405  \pm 0.011$	&$0.17   \pm 0.010$	  &$0.095 \pm 0.010$\\ 
COD	&0.58	&0.71	&0.33	&0.14\\
R	  &0.76	&0.84	&0.58	&0.37\\
\hline
Lat &$-33^\circ$               &$-41^\circ$              	&$-51^\circ$                  &$-62^\circ$\\
\hline
y0	&$0.01    \pm 0.008$ &$-0.021 \pm 0.01$	  &$-0.037 \pm 0.02$	&$-0.044 \pm 0.04$\\
xc	&$-52.38  \pm 7.84$	 &$-68.66 \pm 6.09$	  &$-11.64 \pm 3.28$	&$16.12 \pm 1.43$\\
w	  &$146.88  \pm 3.28$	 &$154.26 \pm 2.62$	  &$142.61 \pm 1.56$	&$145.21 \pm 0.75$\\
A	  &$0.172   \pm 0.011$ 	&$0.39  \pm 0.018$	&$0.93   \pm 0.03$	&$2.88 \pm 0.049$\\
COD	&0.33	&0.48	&0.66	&0.87\\
R	  &0.57	&0.69	&0.81	&0.93\\
\hline
\multicolumn{5}{l}{Approximation was made using the sinusoidal function: equation (\ref{fsin}).}\\
\multicolumn{5}{l}{Lat -- heliolatitude, R -- correlation coefficient, COD -- coefficient of determination.}\\
%&The table was calculated using for the approximation the sinusoidal model&\\ 
%$y = y0 + A* sin(pi*(x-xc )/w) $,
%In the table: Lat - heliolatitude, R - correlation coefficient and COD - coefficient of determination ($R^2$).}
 \end{tabular} \\
 % <data>
\end{table}

%%%%%%%%%%%%%%%%%%%%%%%%%%%%%%%%%%%%%Figure 6

For most latitudes, the approximation was performed successfully: of the 18 time profiles shown in Figures~\ref{nappr}, \ref{sappr} in 14 cases there are variations with close values of the period. The exception is latitudes $26^\circ$   in both hemispheres. This mid-latitude interval is characterized by a predominance of short-period variations over the long-period variations which are present at other latitudes. This observation is confirmed by Figure~\ref{TandAmp}, showing the amplitude and period of this long-period variation. While the amplitude of the variation decreases almost monotonously from the poles to the equator, the period of variation experiences a sharp drop at latitudes located within the zone of sunspot  formation. Both at higher latitudes and after this drop the period of variation remains close to its  average value (average half-period equals to $w = 149 \pm 1$ Carrington rotations). This time interval corresponds to a period of 22.3 years, i.e., to the typical duration of the Sun's magnetic cycle (the Hale cycle).

%%%%%%%%%%%%%%%%%%%%%%%%%%%%%%%%%%%%%Figure 7

The calculation of the correlation coefficient (Figure~\ref{corr}) allows  to evaluate the quality of the approximation of the field changes by the sinusoidal function.
The high values of the correlation coefficient between the experimental data and approximating sinusoidal function  give evidence of the good correspondence of the two data sets.
At all latitudes except the intervals $26^\circ$ and  $33^\circ$   of the northern and southern hemispheres, the correlation coefficient $R>0.6$. The largest values  of the $R$ coefficient 
are reached at high latitudes ($R = 0.91$ for the northern hemisphere and $R= 0.93$ for the southern hemisphere), where the dipole nature of the magnetic field is clearly visible,
and fluctuations are significantly less than the regular component 
of the magnetic field (Figure~\ref{cols}a and \ref{tabs}i). A sharp decline in correlation occurs in the latitude interval of $26^\circ$  and $33^\circ$. It is quite natural that the dipole 
component of the field is most clearly manifested at near-polar latitudes. More surprising is the fact, that high coefficient values ($R> 0.8$) were also observed near equator at low latitudes -- 
about $+10^\circ$  and $-10^\circ$.

\section{Conclusions}
Variations of the weak magnetic fields of the photosphere over the period from 1978 to 2016 were examined. Time-latitude diagram for this time interval was constructed and a set of time profiles of the magnetic field was selected for the further analysis. The set consisted of 18 time profiles of the magnetic field at latitudes distributed evenly in the time-latitude diagram from the north to south pole. To highlight long-periodic changes, the curves were smoothed by the sliding smoothing over 21 Carrington rotations. The approximation by sinusoidal function showed that the main period of variations of the magnetic field at different latitudes coincides with the 22-year magnetic cycle of the Sun. The period of this variation is almost constant for all latitudes and averages 22.3 years, a typical duration of the Sun's magnetic cycle (the Hale cycle). This variation  exists at all latitudes except latitudes $26^\circ$  and 33$^\circ$  in the northern hemisphere and $26^\circ$   in the southern hemisphere. These mid-latitude intervals are characterized by a predominance of short-period variations. The amplitude of the 22-year variation decreases almost monotonously from the poles to the equator.

The high values of the correlation coefficient $R$ show that the field changes are successfully approximated by sinusoidal function. In all cases $R$ was above 0.6, except for the latitudes $26^\circ$ and $33^\circ$   in the northern hemisphere and $26^\circ$   in the southern hemisphere, where there is a sharp decline of correlation. The largest values of the $R$ coefficient were reached at high latitudes ($R = 0.91$ for the northern hemisphere and $R = 0.93$ for the southern hemisphere), where the dipole nature of the magnetic field is clearly visible. High values of the coefficient 
($R >0.8$) were also observed near the equator at latitudes of about $+10^\circ$  and $-10^\circ$. These results show that variations in the magnetic field with a period of 22 years are a characteristic feature present at most of the heliolatitudes.

\section{Acknowledgements}
The authors thank the reviewer for carefully reading the article and helpful remarks. The NSO
/Kitt Peak data used here are produced cooperatively by NSF/NOAO,NASA/GSFC, and NO-AA/SEL (ftp://nispdata.nso.edu/kpvt
/synoptic/mag/). Data acquired by SOLIS instruments were operated by NISP/NSO/AURA/NSF (https://magmap.nso.edu/solis
/archive.html

\end{document}